# Crop Knowledge Discovery Based on Agricultural Big Data Integration

Vuong M. Ngo
School of Computer Science, University College Dublin
Belfield, Dublin 4, Ireland
vuong.ngo@ucd.ie

M-Tahar Kechadi
School of Computer Science, University College Dublin
Belfield, Dublin 4, Ireland
tahar.kechadi@ucd.ie

## ABSTRACT

Nowadays, the agricultural data can be generated through various sources, such as: Internet of Thing (IoT), sensors, satellites, weather stations, robots, farm equipment, agricultural laboratories, farmers, government agencies and agribusinesses. The analysis of this big data enables farmers, companies and agronomists to extract high business and scientific knowledge, improving their operational processes and product quality. However, before analysing this data, different data sources need to be normalised, homogenised and integrated into a unified data representation. In this paper, we propose an agricultural data integration method using a constellation schema which is designed to be flexible enough to incorporate other datasets and big data models. We also apply some methods to extract knowledge with the view to improve crop yield; these include finding suitable quantities of soil properties, herbicides and insecticides for both increasing crop yield and protecting the environment.

## 1 INTRODUCTION

Although, annual world cereal production in 2018 and 2019 were 2, 595 million tons [5] and 2, 706 million tons, respectively [6], there are about 124 million people in 51 countries facing food crisis and food insecurity [7]. This will get even worse in the coming years. It is expected that the world population will increase from 7.7 billion in 2019 to 8.5 billion in 2030 [25]. Moreover, with limitation of available freshwater and cropland, crop yields must be significantly increased to satisfy the growing world population by using new farming approaches, such as precision farming also called precision agriculture.

Precision farming uses historical data along with data mining algorithms to make specific decisions for determining which crops and which nutrients with suitable quantities will produce the best crop yield. For examples, instead of applying the same large quantities of pesticides to all crops, you can apply smaller amounts to specific plants. This will certainly reduce the production costs and waste, avoid damaging the environment, and reduce negative effect on some other insects species. The collected historical big data will be mined and analysed so that the whole ecosystem will be taken into account and some key decisions will come out to efficiently use the land and other required resources.

In this paper, we propose a constellation schema which includes many fact and dimension tables containing information about crops, fields, products, operations, testings and management of the farms. This schema is designed to be flexible to integrate any agri datasets in a unified representation. Besides, this schema also can be used to build a data warehouse (DW), adapting quality criteria of agricultural Big Data. In addition, the datasets originated from various sources are extracted, transferred and loaded using the unified schema to become a unified dataset. Intelligent methods based on a range of crops and their crop yields are applied to discover new knowledge and get better understanding of some farming processes. We limit the scope of this paper to discovering knowledge about soil properties, herbicides and insecticides.

The rest of this paper is organised as follows: in the next Section, we reviewed the related work on data integration and knowledge discovery in agriculture. In Section 3, benefits, challenges and proposed schema for agricultural data integration are presented. Section 4 presents a methodology of how to find appropriate quantities of soil properties, herbicides and insecticides for a range of crops, based on Big Data. Finally, we conclude and comment on future work in Section 5.

## 2 RELATED WORK

There are numerous research works in the literature about the analysis and mining of agricultural datasets with the view to improve farming operations. These include crop yield increments, pest control, early warning, and farm management. In [13], data mining techniques were applied on crop, soil and climatic datasets to maximise the crop production. While, [4] combined surveys of farming practices with model-based simulations to determine the relation between weeds and crop yield. In [22], the authors predicted pest population dynamics using time series clustering and structural change detection of different pest species and groups. In [24], the authors provide optimal management solutions to efficiently identify nutrients and water; a multi-objective genetic algorithm was used to implement an E-Water system. Finally, in [19], [20] and [21] the authors presented interesting decision support systems for early warning, soil nutrient and financial services, respectively. However, all the mentioned works did not tackle agricultural Big Data integration. So, their datasets contain only a reasonably small agricultural information.

Data integration is very important task in large enterprises and organisations, which own various data sources. It is implemented in large-scale scientific projects. Without it, it is very challenging to access data across many autonomous and heterogeneous data sources and formats [9]. The paper [1] proposed a set of ontologies to facilitate the agricultural data integration, the authors in [23] proposed to loaded the data into RDF triples before being transformed into a relational schema. In [2], the authors proposed a system based on JSON format and provided an API, which centralised and standardised to communicate with different applications to integrate the various datasets. They chose PostgreSQL being a relational database, as the database management system. Further, in [8], a relational schema and geographic information were used to build a decision support system about weather, crops, regions and bugs. However, the ontologies and relational schemas are not flexible for adding new datasets and cannot deal with high-performance data analysis. To overcome these limitations, [15] and [16] used data warehouse constellation schema to integrate agricultural Big Data. However, their schema does not include some important information about farming operation, such as crop, soil and nutrient testings, and treatment, spray, fertiliser and zone managements. Besides, they are about data warehouse design and implementation, it did not use data mining algorithms to discovery crop knowledge.

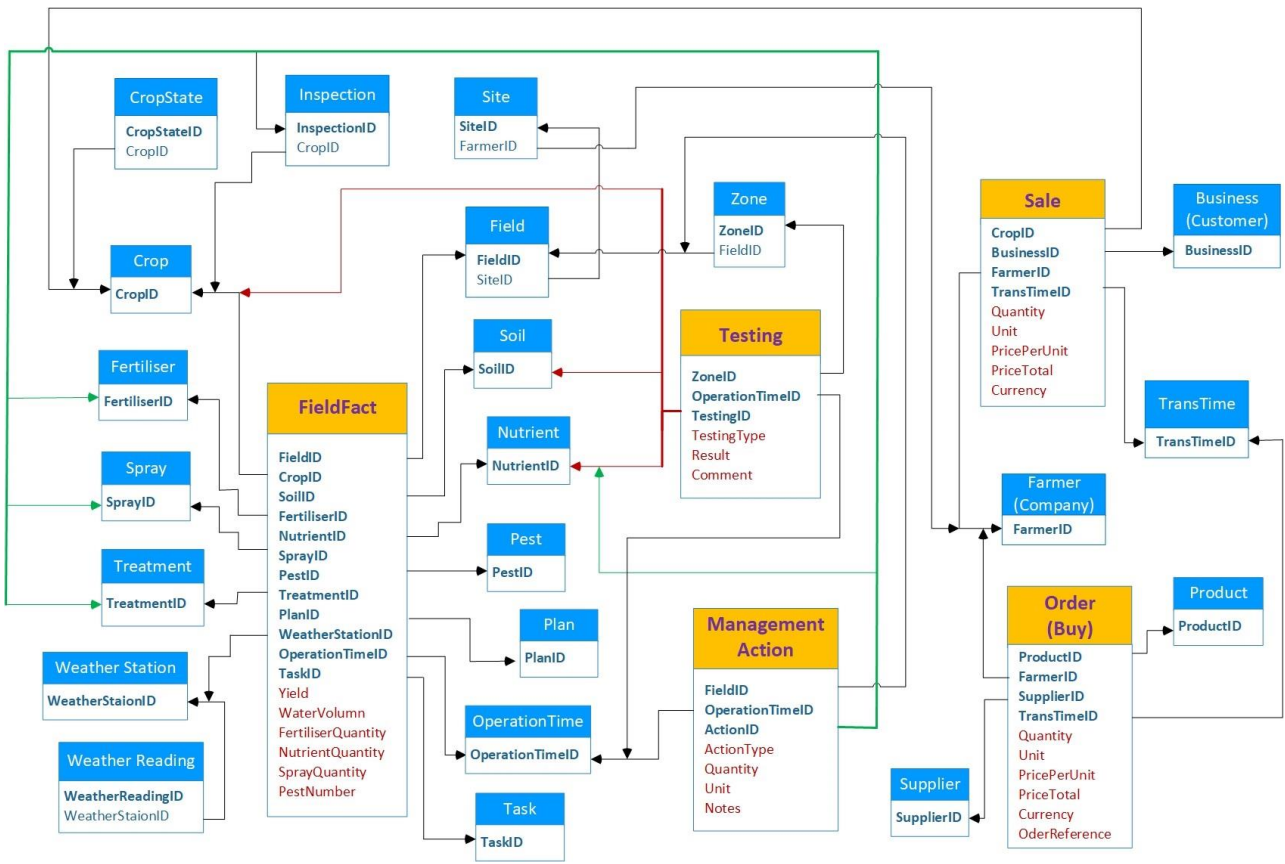

**Figure 1: A part of our constellation schema for Precision Agriculture**

## 3 AGRICULTURAL BIG DATA SCHEMA

Integrating Agricultural data from various sources would produce estimates that can be used nationwide and other specialised organisations. These estimates include supply and demand, farming income, gross domestic products, etc. Besides, data integration not only reduces both costs and burden on survey respondents, but also leads to higher quality information. High quality integrated data will increase the quality of the mined results and therefore, help farmers manage efficiently their operations and make bet- ter decisions to fulfil specific needs, such as optimising fertiliser, insecticides, irrigation, etc.

Raw and semi-processed agricultural data are not only collected through various sources, but also very large, complex, unstructured, heterogeneous, non-standardised, and inconsistent. Specifically, the agricultural data has all the features of Big Data: (1) Volume: The amount of agricultural data is rapidly increasing and is intensively produced by endogenous and exogenous sources, such as operation processes, sensors, satellites, farm equipment, government agencies, retail agronomists, seed companies, and farmers. The exogenous sources can help supply information about local pest and disease outbreak tracking, market accessing, food security, products and prices; (2) Variety: Agricultural data has many different types and formats; structured and unstructured data, video, imagery, chart, metrics, geo-spatial, multimedia, models, equation and text; (3) Velocity: The produced and collected data increases at high rates, as sensing and mobile devices are becoming more efficient and cheaper. The datasets need to be cleaned, aggregated and harmonised in real-time; (4) Veracity: The tendency of agronomic data is uncertain, inconsistent, ambiguous and error prone because the data is gathered from heterogeneous sources, sensors and manual processes. So, agricultural Big Data integration is very challenging.

**Table 1: Descriptions of other dimension tables**

| No. | Tables | Particular attributes |
|---|---|---|
| 1 | Business | BusinessID, Name, Address, Phone, Email |
| 2 | Crop | CropID, CropName, VarietyID, VarietyName, EstYield, SeasonStart, SeasonEnd, BbchScale, Scien.Name, HarvestEquipment, Equ.Weight |
| 3 | Crop State | CropStateID, CropID, StageScale, Height, MajorStage, MinStage, MaxStage, Diameter, AveHeight, CoveragePercent |
| 4 | Farmer | FarmerID, Name, Address, Phone, Mobile, Email |
| 5 | Fertiliser | FertiliserID, Name, Unit, Status, Description, GroupName |
| 6 | Field | FieldID, FieldName, SiteID, Reference, Block, Area, AreaUnit, WorkingArea, WorkingAreaUnit, Latitude, Longitude, Geometric-Points, FieldImage, Notes |
| 7 | Inspection | InspectionID, CropID, Description, PestType, Severity, AreaValue, AreaUnit, Order, Date, Notes, GrowthStage |
| 8 | Nutrient | NutrientID, NutrientName, Date, Quantity |
| 9 | Operation Time | OperationTimeID, StartDate, EndDate, Season |
| 10 | Pest | PestID, CommonName, ScientificName, PestType, Description, Density, MinStage, MaxStage, Coverage, CoverageUnit |
| 11 | Plan | PlanID, PlanName, RegisNo, ProductName, ProductRate, Date, WaterVolume |
| 12 | Product | ProductID, ProductName, GroupName |
| 13 | Spray | SprayID, SprayProductName, ProductRate, Area, WaterVolume, ConfDuration, ConfWindSpeed, ConfDirection, ConfHumidity, ConfTemp, ActivityType |
| 14 | Site | SiteID, FarmerID, SiteName, Reference, Address, GPS, CreatedBy |
| 15 | Soil | SoilID, NutrientID, PH, Nitrogen, Phosphorus, Potassium, Magnesium, Calcium, CEC, Silt, Clay, Sand, SoilTexture, SoilType, Organic-Matter, TopSoil, SupSoil, TestDate, Unit |
| 16 | Supplier | SupplierID, SupplierName, Address, Phone, Email |
| 17 | Task | TaskID, Desc, Status, TaskDate, TaskInterval, CompDate, AppCode |
| 18 | TransTime | TransTimeID, OrderDate, DeliverDate, ReceivedDate |
| 19 | Treatment | TreatmentID, TreatmentName, FormType, LotCode, Rate, AppCode, LevlNo, Type, Description, AppDesc, TreatmentComment |
| 20 | Weather Reading | WeatherReadingID, WeatherStationID, ReadingDate, ReadingTime, AirTemper, Rainfall, SPLite, RelativeHumidity, WindSpeed, Wind-Direction, SoilTemper, LeafWetness |
| 21 | Weather Station | WeatherStationID, Station Name, Latitude, Longitude, Region |
| 22 | Zone | ZoneID, ZoneName, FieldID, SoilID, ZoneType, Area, AreaUnit, Latitude, Longitude, GeometricPoints, YieldMap, SatellitePicture, Notes |

Data integration is a data combination and validation collected from different sources. It provides a user with a unified view of the whole data [12]. The process of integrating data uses a schema, similar to databases (DB), in which data coming from different sources can be extracted, assessed, validated, and organised following its metadata model. Unlike DB schema, a DW schema is a collection of objects, including tables, views, indexes, and synonyms which consists of some fact and dimension tables [17]. There are three kind of DW schema models; namely star, snowflake and constellation [11]. Agriculture DW is an enterprise data warehouse, and requires a number of fact tables (or subjects or views). It is usually represented as fact constellation schema.

We developed an agricultural fact constellation schema using all the information collected the original data sources (operational databases) and the requirements of farmers, companies, and agronomists. Figure 1 gives an overview of the proposed DW schema. It includes five fact tables being FieldFact, Sale, Order, Testing and Management Action, and 22 dimension tables. The FieldFact fact table, which has 12 dimensions and six measures, contains information about fields, soil, fertiliser, nutrient, weather, treatment, and pest. While, the Order and Sale fact tables contain data about farmers' trading operations which have the same four dimensions, fertiliser, treatment, inspection and spray through ActionID and ActionType attributes.

Each dimension table contains details about instances of an object involved in a crop yield and farm management. Table 1 describes the main attributes of 22 dimension tables. For examples, the Field table contains information about fields, such as name, area, longi- tude, latitude and geometric information. A field often has many zones. The Zone table contains specific information on every zone; zone type, soil and yield map on zone. The Soil table describes soil properties, such as pH value, nitrogen, phosphorus, texture and organic matter.

# 4   CROP KNOWLEDGE DISCOVERY

## 4.1   Datasets

The datasets which are used to validate the proposed fact constellation schema were obtained from a leading commercial agronomy service company in the United Kingdom. The company collected data from its operational systems, research results, and field trials. Especially, they collected real agricultural data in iFarms, B2B sites, technology centres and demonstration farms at Belgium, Brazil, Ireland, Poland, Romania, Ukraine and United Kingdom [18]. They have 800 sale forces, 112 distribution points, 34 input formulation and processing facilities, 73 demonstration farms, 12 million hectares of direct farm customer footprints and 50, 000 trial units.

There is a total of 29 datasets. On average, each dataset contains and have six and five measures, respectively. The Testing fact table presents testing operation about crop, soil and nutrient through TestingID and TestingType attributes. Finally, the Management Action fact table describes management operations about nutrient, 18 tables stored in 1.4 GB storage size. Each dataset focuses on a few information of farming operations. For example, crop dataset almost contains information about crop, such as name, estimated yield, harvest equipment, BBCH growth stage index, major stage, diameter and crop coverage percent. While, the weather dataset includes information on location of weather station, air and soil temperature, rainfall, humidity and wind speed direction over time. In pest dataset, there is information about name, type, density, stage, coverage and detected date of pests.

## 4.1   Crop Yield Classification

In every field, we extract information related to crop yield from the new schema; field identification, year, season, crop name, yield, soil properties (i.e. soil pH, soil P, soil K, soil Mg), herbicides and insecticides. Each record, based on crop type and yield, can be classified into one of the five groups. Every group contains 20% of the number of records of each crop type. Group 1 is the highest 20% yield group and Group 5 is the lowest with 20% yield group in every crop type.

The Table 2 describes the 12 most popular crops in the EU countries. These are Spring Barley (Barley S.), Winter Barley (Barley W.), Spring Dried Beans (Beans S.), Winter Dried Beans (Beans W.), Grass, Spring Linseed (Linseed S.), Forage Maize (Maize F.), Winter Oats (Oats W.), Winter Rape (Rape W.), Winter Rye (Rye W.), Spring Wheat (Wheat S.) and Winter Wheat (Wheat W.). For each crop type, the mean yield of each group and different percentages be- tween groups are also presented in this table. For examples, Spring Barley belonging to group 1 has mean yield of 8.93 ton/ha and higher than corresponding medium group (group 3) about 36.9%. While, the mean yield of Spring Barley group 5 has only 4.26 ton/ha

**Table 2: Descriptions of mean crop yield (ton/ha) in every yield group**

| Group | Crop | M. Yield | % | Crop | M. Yield | % |
|---|---|---|---|---|---|---|
| 1 | Barley S. | 8.93 | +36.9 | Barley W. | 13.04 | +78.7 |
| 2 | Barley S. | 7.32 | +12.2 | Barley W. | 8.29 | +13.7 |
| 3 | Barley S. | 6.52 | 0 | Barley W. | 7.30 | 0 |
| 4 | Barley S. | 5.81 | -10.9 | Barley W. | 6.40 | -12.2 |
| 5 | Barley S. | 4.26 | -34.8 | Barley W. | 5.16 | -29.3 |
| 1 | Beans S. | 5.21 | +37.3 | Beans W. | 6.15 | +23.6 |
| 2 | Beans S. | 4.32 | +13.9 | Beans W. | 5.51 | +10.8 |
| 3 | Beans S. | 3.79 | 0 | Beans W. | 4.97 | 0 |
| 4 | Beans S. | 1.92 | -49.3 | Beans W. | 4.52 | -9.2 |
| 5 | Beans S. | 1.08 | -71.4 | Beans W. | 3.40 | -31.7 |
| 1 | Grass | 23.80 | +67.7 | Linseed S. | 2.28 | +75.5 |
| 2 | Grass | 21.73 | +53.1 | Linseed S. | 1.57 | +20.9 |
| 3 | Grass | 14.19 | 0 | Linseed S. | 1.30 | 0 |
| 4 | Grass | 9.01 | -36.5 | Linseed S. | 0.84 | -35.7 |
| 5 | Grass | 7.62 | -46.3 | Linseed S. | 0.43 | -67.1 |
| 1 | Maize F. | 47.00 | +16.7 | Oats W. | 8.06 | +15.1 |
| 2 | Maize F. | 44.67 | +10.9 | Oats W. | 7.50 | +7.1 |
| 3 | Maize F. | 40.27 | 0 | Oats W. | 7.00 | 0 |
| 4 | Maize F. | 32.63 | -19 | Oats W. | 6.93 | -1 |
| 5 | Maize F. | 21.62 | -46.3 | Oats W. | 5.64 | -19.4 |
| 1 | Rape W. | 4.59 | +27.7 | Rye W. | 39.90 | +41.4 |
| 2 | Rape W. | 4.00 | +11.4 | Rye W. | 32.39 | +14.7 |
| 3 | Rape W. | 3.59 | 0 | Rye W. | 28.23 | 0 |
| 4 | Rape W. | 3.15 | -12.5 | Rye W. | 23.19 | -17.8 |
| 5 | Rape W. | 2.36 | -34.3 | Rye W. | 17.77 | -37 |
| 1 | Wheat S. | 7.20 | +27.9 | Wheat W. | 11.74 | +25.9 |
| 2 | Wheat S. | 6.52 | +15.8 | Wheat W. | 10.22 | +9.6 |
| 3 | Wheat S. | 5.63 | 0 | Wheat W. | 9.32 | 0 |
| 4 | Wheat S. | 4.73 | -16 | Wheat W. | 8.55 | -8.3 |
| 5 | Wheat S. | 1.94 | -65.6 | Wheat W. | 6.83 | -26.7 |

and lower than its corresponding medium group which is about 34.8%.

## 4.2 Crop and Soil Correlation

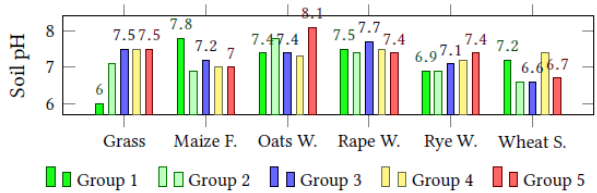

**Figure 2: Soil pH values**

Soil acidity (pH) holds an important role in soil fertility. Maintaining the soil pH at a suitable level will certainly increase the microbiological activity and nutrient of the soil. Some crops often grow best in a pH range 6.0 to 7.0 , others grow well under slightly acidic conditions. The mean soil pH values for six crops classified by yield groups are shown in Figure 2. From this figure, we can detect the desirable soil pH value for optimum growth to make the best yield for certain crops.

There are clearly differences between soil pH values among yield groups in Grass, Forage Maize and Winter Rye. While, Spring Wheat has small difference, Winter Oats and Winter Rape do not show any difference among yield groups. So, we did not find optimal soil pH values for Winter Oats and Winter Rape. The optimal soil pH values for Grass, Forage Maize, Winter Rye and Spring Wheat are 6.0, 7.8, 6.9 and 7.2, respectively.

Moreover, the supplied soil nutrients are also critical for crop growth. The largest amounts of soil nutrients required by crops are phosphorus (P), potassium (K) and magnesium (Mg). They are often considered as the most important nutrients. The mean soil P,

K and Mg quantities for crops classified by yield groups are shown in Figures 3, 4 and 5, respectively.

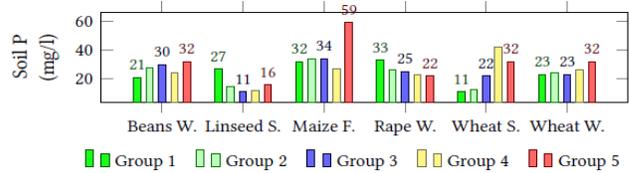

**Figure 3: Mean soil P quantities**

In Figure 3, the mean soil P quantities are clearly different among yield groups in Spring Linseed, Winter Rape, Spring Wheat and Winter Wheat. While, in Winter Dried Beans and Forage Maize, the differences are not very clear, but they are enough to determine suitable P quantities. The optimal P quantities are 21 (mg/l) for Winter Bean, 27 (mg/l) for Spring Linseed, 32 (mg/l) for Forage Maize, 33 (mg/l) for Winter Rape, 11 (mg/l) for Spring Wheat and 22 (mg/l) for Winter Wheat.

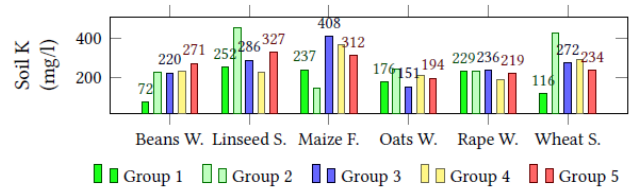

**Figure 4: Mean soil K quantities**

In Figure 4, the differences among yield groups in Spring Linseed, Winter Oats and Winter Rap are not significant. So, we did not find optimal K quantities for these crops. Meanwhile, for the other crops the difference is significant and we can find optimal soil K quantities; which are 72 (mg/l) for Winter Dried Beans, 237 (mg/l) for Forage Maize, and 116(mg/l) for Spring Wheat.

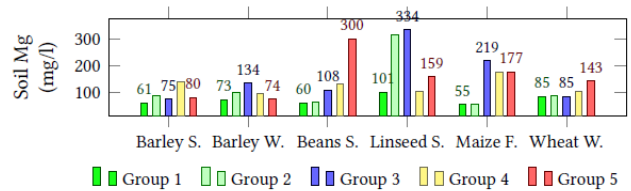

**Figure 5: Mean soil Mg quantities**

In Figure 5, there is significant differences among yield groups, the mined optimal Mg quantities for Spring Barley, Spring Dried Beans, Forage Maize and Winter Wheat are 61 (mg/l) , 60 (mg/l), 55 (mg/l), and 85 (mg/l), respectively. However, for Winter Barley there is no significant difference between group 1 and group 5, and for Spring Linseed, there is no significant difference between group 1 and group 4. So, we could not find optimal Mg quantities.

## 4.4 Crop and Herbicides Correlation

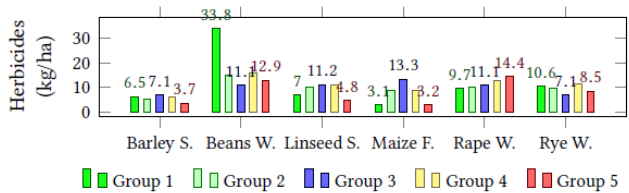

**Figure 6: Mean herbicide quantities**

Weed control is generally considered to be essential for crop growth and it is often dealt with herbicides. However, the use of herbicides must be reduced because they have negative effects on the environment and create very bad and health issues. The key question then is to see whether herbicide increments will lead to an increase in crop yield or not. Figure 6 presents mean herbicide quantities used in every yield group of some crops. In almost all crops; Spring Barley, Spring Linseed, Forage Maize, Winter Rape and Winter Rye, herbicide increments do not increase crop yield, in some cases, they even reduce the yield. With an exception to the rule, for Winter Bean, the highest yield group used the highest herbicide quantity, which is about 33.8 (kg/ha).

## 4.5 Crop and Insecticides Correlation

The use of insecticides to keep away bugs and pests does not only cause economic and health issues, but also it directly affects other insects and animals, such as bee, beetles, frogs, etc. So, for every crop, one needs to determine suitable insecticide quantities.

The optimal insecticide quantities in yield groups of Spring Barley, Winter Barley, Spring Dried Beans, Spring Linseed, Winter Rape and Winter Rye are shown in Figure 7. There are significant differences among yield groups, so the optimal insecticide quantities that have been mined are 736 (g/ha) for Spring Dried Beans, 693 (g/ha) for Winter Rape and only 79 (g/ha) for Winter Rye. While, the Spring Barley, Winter Barley and Spring Linseed did not show significant differences between high yield groups and low yield groups. So, we could not mine optimal insecticide quantities for these crops.

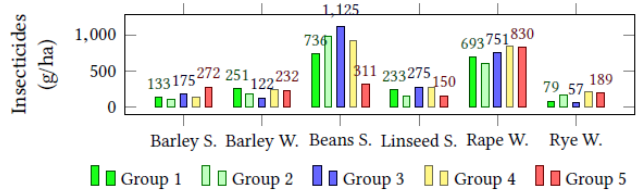

**Figure 7: Mean insecticide quantities**

## 5 CONCLUSION AND FUTURE WORK

In this paper, we proposed a fact constellation schema as an overall framework for integrating various agricultural datasets. The schema is flexible and extensible to other agricultural datasets and quality criteria of Big Data analytics. Based on this schema, we extracted, transferred and loaded information from various datasets into a unified representation of agricultural dataset. We also presented a data analytics study about the effect of certain agriculture factors on crop yields using classification techniques. We even find optimal quantities of soil properties, herbicides and insecticides for certain crops under the study.

As a future work, we will apply more sophisticated machine learning algorithms on our unified dataset to discover global relations between soil properties together and other factors, such as nutrients and fertilisers. This future study will be supported by an intelligent visualisation interface, graph representation [10] and ontology [3], [14] for accessing data access and showing the results of the data analysis.

## ACKNOWLEDGMENT

This research is funded under the SFI Strategic Partnerships Programme (16/SPP/3296) and is co-funded by Origin Enterprises Plc.